\documentclass[aps,prl,twocolumn,floatfix,showpacs]{revtex4}

\usepackage{epsfig}
\usepackage{amsmath}

\begin{document}

\title{The absorption spectrum around $\nu =1$: evidence for a small size
Skyrmion}
\author{J. G. \surname{Groshaus}}
\email{j.groshaus@weizmann.ac.il}
\author{V. \surname{Umansky}}
\author{H. \surname{Shtrikman}}
\author{Y. \surname{Levinson}}
\author{I. \surname{Bar-Joseph}}
\affiliation{Department of Condensed Matter Physics, The Weizmann
Institute of Science, Rehovot, Israel.}


\begin{abstract}
We measure the absorption spectrum of a two-dimensional electron
system (2DES) in a GaAs quantum well in the presence of a
perpendicular magnetic field. We focus on the absorption spectrum
into the lowest Landau Level around $\nu =1$. We find that the
spectrum consists of bound electron-hole complexes, trion and
exciton like. We show that their oscillator strength is a powerful
probe of the 2DES spatial correlations. We find that near $\nu =1$
the 2DES ground state consists of Skyrmions of small size (a few
magnetic lengths).
\end{abstract}

\maketitle

Electron-electron interactions are known to play an important role
in determining the ground state of a two-dimensional electron
system (2DES) around $\nu =1$. They favor a ferromagnetic order at
this filling factor even in the absence of a Zeeman spin gap.
Furthermore, it was suggested that if this spin gap is
sufficiently small, the ground state around $\nu =1$ consists of
charged spin-textures excitations known as Skyrmions \cite
{Sondhi93, Fertig94}: neighboring electrons tend to align their
spin in parallel to minimize the exchange energy, creating a
smooth texture of the spin field, in which many electrons
participate. The projection of the spin along the magnetic field
direction, $S_{z}$, varies gradually from the center of the
texture towards the edges, with the total integrated charge
being $-|e|$ for $\nu >1$ (Skyrmion) and $+|e|$ (anti-Skyrmion) for $\nu <1$%
. It has been realized that as the spin gap increases such a large
size structure becomes very costly in energy. Less electrons find
it energetically favorable to flip their spin, and the Skyrmion
size shrinks. Theoretical estimates for GaAs ($g=-0.44$) set the
Skyrmion size at $B=8$ T to be about two magnetic lengths only
\cite{Fertig94}.$\ $The experimental evidence for the Skyrmion
theory comes primarily from NMR measurements \cite{Barrett95},
tilted magnetic field transport experiments \cite{Schmeller95},
and optical absorption spectroscopy \cite{Aifer96}. In these
measurements the spin polarization $\langle S_{z}\rangle $ of the
2DES was determined, and was shown to fall abruptly at both sides
of $\nu =1$, as predicted by the theory.

In this paper we wish to use optical absorption spectroscopy to
study the 2DES ground state in this regime. A major problem in
interpreting optical measurements of 2DES is how to account for
the strong Coulomb interaction between the valence hole and the
surrounding electrons. Recent experimental and theoretical works
show that the optical spectrum in the fractional quantum Hall
regime should be understood in terms of bound electron-hole
complexes, e.g. neutral and charged (trions) excitons
\cite{Yusa2001,Cooper97,Wojs,Schuller2003}. In this paper we find
clear evidence for the formation of these bound electron-hole complexes around $%
\nu =1$ and show that they can be a powerful probe for the ground
state of the 2DES. A measurement of their oscillator strength (OS)
and energy dependence on $\nu $ allows us to estimate the Skyrmion
size, and we show that only a few electrons significantly
contribute to it.

The experiments were done at a temperature of 4.2 K, and a
magnetic field of up to 9 T applied along the growth axis of the
wafer. The light source was a tunable Ti-sapphire ring laser, and
the sample was illuminated with power densities $I_L$ lower than 1
mW/cm$^{2}$ through a thick plastic optical fiber and a circular
polarizer. Reversing the polarization of the light was
accomplished by reversing the direction of the magnetic field $B$.
To obtain the absorption spectrum we measured the photo-current
(PC) flowing between the 2DES and a back gate. We have previously
demonstrated this technique in obtaining the absorption spectrum of a 2DES at $B=0$ using a front gate\cite%
{Yusa2000}. We find that the PC at laser energies below the fundamental gap
is negligible, resulting in background free spectral measurements. This also
gives us confidence that the measured PC spectrum indeed reflects the
quantum well absorption spectrum. This was further verified by conducting
complementary photoluminescence excitation measurements, where similar
qualitative behavior of the peak positions and line shape were observed.
Several samples with the same general structure were investigated, all
consisting of a single 20 nm GaAs/Al$_{0.3}$Ga$_{0.7}$As modulation-doped
quantum well grown on top of a 0.5 $\mu $m Al$_{0.3}$Ga$_{0.7}$As barrier
layer, separating it from the back gate layer. The wafers were processed to
a mesa structure with selective ohmic contacts to the 2DES and to the
back-gate. Applying a voltage between the 2DES and the back gate we could
tune the electron density $n_{e}$ continuously.

\begin{figure}[tbp]
\vspace*{3mm} {\hbox{\epsfxsize=83mm \epsffile{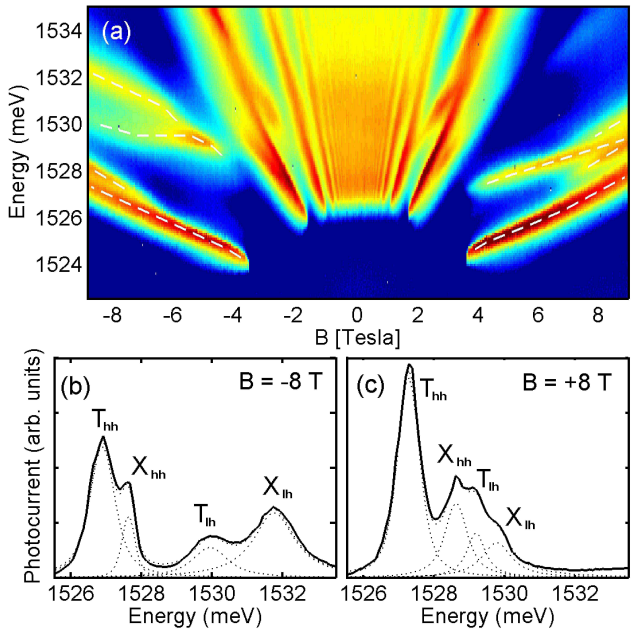} }}
\caption{(a) (color) Photocurrent spectra for $n_{e}=1.7\times
10^{11}$ cm$^{-2}$ and $I_L=0.3$ mW/cm$^2$. The dashed lines are
guide to the eye. (b, c) Spectra at both polarizations (solid
line) for $\protect\nu <1$, where both the singlet trion $T$ and
the exciton $X$ are present for heavy and light holes. The dotted
lines are the fitted functions.} \label{fig1}
\end{figure}

Figure 1(a) presents a compilation of PC measurements at constant
$n_{e}$ as $B$ is varied between $-9$ T and $+9$ T. The magnitude
of the PC is color-coded, with dark-red color indicating strong
absorption. The characteristic Landau levels fan is well resolved
even at low fields. The lowest energy line for both signs of $B$
corresponds to an absorption process from the heavy-hole (hh) band
into the lowest electron Landau level (LLL). It has a sharp onset
at $B=\pm 3.6$ T, corresponding to $\nu =2$: below this magnetic
field the
LLL is full, so that no absorption is possible. This allows us to obtain $%
n_{e}$ at each gate voltage. A few meV above this line and parallel to it
there is an additional line in both directions of $B$. It corresponds to the
creation of a light-hole (lh) - electron pair. We note that the hh and lh
lines (marked by dashed lines) at a given direction of $B$ probe different
electronic Zeeman levels: for $B<0$ a $\sigma ^{+}$ photon tuned to the hh
transition creates a valence hole with $J_{z}=3/2$\ and an electron with $%
S_{z}=-1/2$. When tuned to the lh transition it creates a valence hole with $%
J_{z}=1/2$ and an electron with $S_{z}=1/2$. In other words, for
$B<0$ the hh transition probes the electron lower Zeeman state
$|\uparrow\rangle$, while the lh probes the upper Zeeman state
$|\downarrow\rangle$. Similarly, for $B>0$ the hh and lh
transitions create an electron in the $|\downarrow \rangle$ and
$|\uparrow\rangle$ states, respectively.

Examining Fig. 1(a) carefully we note that at high magnetic fields, below $%
\nu =1$, a new absorption line appears above each of the lh and hh LLL
lines. These new lines are particularly visible at the negative fields \ ($%
B<0$) spectra, where the large hh-lh splitting allows us to easily
resolve them. Figures 1(b) and 1(c) show the spectrum at $B=-8$ T
and $+8$ T, respectively, above the branching field. It is evident
that the spectrum consists of four peaks at each light
polarization.

To identify the nature of the peaks we followed their evolution
with density. We found that as the electron density is decreased,
for both hh and lh transitions the lower energy peak evolves
continuously into the singlet trion $T$ while the high energy peak
developes into the neutral exciton $X$ \cite{CommentTriplet}.
These bound complexes were intensively studied in dilute 2DES
during the last decade, and their identification is well
established \cite{Yusa2000,Finkelstein95}. We note that such
continuous evolution of the $X $ and $T$ peaks from the dilute
limit to higher electron densities was observed in both
photoluminescence and absorption measurements in the fractional
quantum Hall regime \cite{Yusa2001, Schuller2003}. We therefore
conclude that the lines that are observed around $\nu =1$ are due
to the formation of bound complexes of a similar nature.

\begin{figure}[tbp]
\vspace*{3mm} {\hbox{\epsfxsize=83mm \epsffile{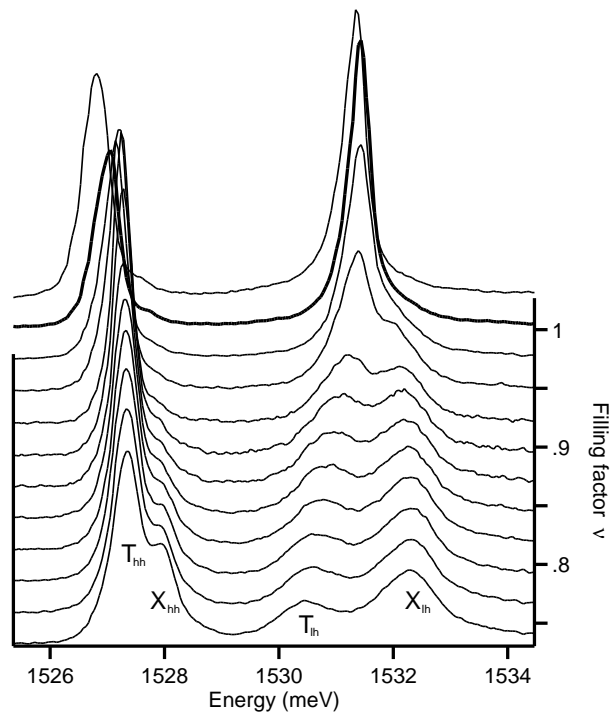} }}
\caption{Evolution of the spectra of both heavy and light hole
excitons $X$ and trions $T$ as a function of the density at $B=-9$
T and $I_L=3$ $\mu$W/cm$^2$ . The $\protect\nu =1 $ spectrum is
emphasized by a thick line. It is seen that both $X$ lines appear
abruptly below $\protect\nu =1$.} \label{fig2}
\end{figure}

Figure 2 shows the evolution of spectrum as $n_{e}$ is decreased below $\nu
=1$ \cite{Comment-ne}. It is clearly seen that at $\nu =1$ new satellite
peaks, $X_{hh}$ and $X_{lh}$, abruptly appear above the two lower energy
peaks, $T_{hh}$ and\ $T_{lh}$, and gain OS as the density is further
reduced. We repeated these measurements in several samples at various
magnetic fields and electron densities and confirmed that this is indeed a
filling-factor dependent phenomena that occurs at $\nu =1$ for both hh and
lh transitions and both light polarizations.

\begin{figure}[tbp]
\vspace*{3mm} {\hbox{\epsfxsize=83mm \epsffile{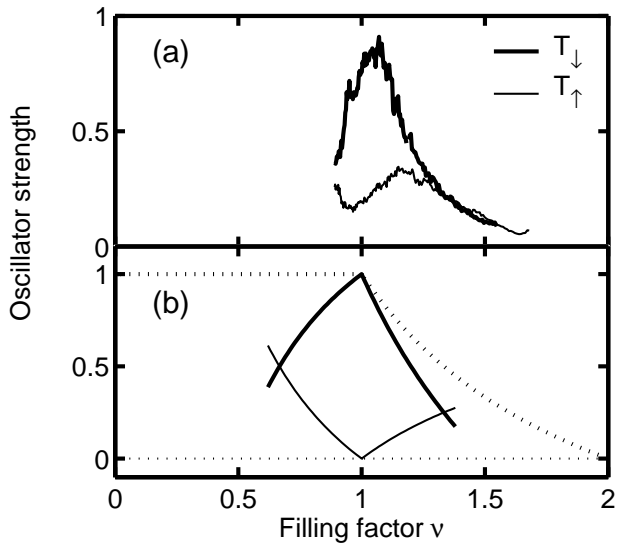} }}
\caption{ (a) The oscillator strength of $T_{lh}$ for both
polarizations at
constant $n_{e}=1.8\times 10^{11}$ cm$^{-2}$. (b) The calculated values of $%
|M_{T,\downarrow }|^{2}$ and $|M_{T,\downarrow }|^{2}$ with $u_{m}=\protect%
\delta _{m,0}$ as $B$ is scanned. The dotted line corresponds to free
electron picture ($u_{m}=0$). The values of $|M|^{2}$ are normalized by
dividing by $N_{\protect\phi }|_{\protect\upsilon =1}$. }
\label{fig3}
\end{figure}

To quantitatively determine the behavior of the $X$ and $T$ peaks
as a function of $\nu $ we have fitted each spectrum, such as
those depicted in Figs. 1(b) and 1(c), to a sum of four Voigt
functions. This allowed us to obtain the parameters of each peak:
the OS, the peak energy and the line width. Figure 3(a) shows the
OS of $T_{lh}$ at both polarizations as $B$ is varied at constant
$n_{e}$. We can see that at $\nu =1$ there is a pronounced maximum
in the absorption into the electron $|\downarrow \rangle $ state
and a corresponding minimum for $|\uparrow \rangle $. Away from
$\nu =1 $ the absorption in both polarizations becomes nearly
equal, indicating a depolarization of the 2DES. We observe a
similar behavior by varying $n_{e}$ at a constant magnetic field.
This dependence of the OS on $\nu $ is similar to that reported by
Aifer \textit{et. al.} \cite{Aifer96}. We note, however, that that
work does not correctly identify the $X$-$T$ splitting and
describes the absorption as a single particle process, neglecting
the electron-hole interaction. It is evident from our data that a
proper interpretation of the experiment should account for the
creation of bound e-h complexes. In the following we present such
a formulation and show that it yields a new insight into the
nature of the ground state of the 2DES.

The wave function of a magneto-exciton in an empty well
($n_{e}=0$) \cite{Cooper97} can be written in the LLL
approximation, which neglects higher Landau levels mixing, as
$|X_{\downarrow }\rangle =\sum_{m=0}^{N_{\phi }-1}h_{m,\uparrow
}^{\dagger }c_{m,\downarrow }^{\dagger }|0\rangle $, where the $%
c_{m,\downarrow }^{\dagger }$ ($h_{m,\uparrow }^{\dagger }$)
operator creates an electron (valence hole) in the corresponding
spin state of the orbital $\phi _{m}(r)$ with angular momentum
$m$, and $N_{\phi }$ is the LLL degeneracy (an analogous
expression can be written for $|X_{\uparrow }\rangle $). On the
other hand, it is well known that upon increasing $n_e$ the
dominant absorption process becomes the creation of a singlet
trion, a bound state of two electrons in a singlet state and a
valence hole. Thus, when dealing with a 2DES near $\nu =1$ we make
the following conjecture: two bound complexes, $X$ and $T$, may be
created in an absorption process, depending on the energy of the
photon. The $X$ complex is created only in orbitals $\phi _{m}$
which are \textit{empty in both spin states}, while $T$ is created
only in orbitals in which \textit{one electron is already
present}. This conjecture preserves the nature of the bound
complexes as in the dilute limit, and in particular - the singlet
character of $T$ \cite{Whittaker97}. The final states upon
absorption into $|\downarrow \rangle $ can thus be written (up to
normalization) as

\begin{eqnarray}
|X_{\downarrow }\rangle  &=&\sum_{m=0}^{N_{\phi }-1}h_{m,\uparrow }^{\dagger
}c_{m,\downarrow }^{\dagger }c_{m,\uparrow }c_{m,\uparrow }^{\dagger }\
|\psi \rangle ,\qquad   \label{final} \\
|T_{\downarrow }\rangle  &=&\sum_{m=0}^{N_{\phi }-1}h_{m,\uparrow }^{\dagger
}c_{m,\downarrow }^{\dagger }c_{m,\uparrow }^{\dagger }c_{m,\uparrow }\
|\psi \rangle ,  \notag
\end{eqnarray}%
where $|\psi \rangle $ is the state of the 2DES prior to absorption, and
analogous expressions can be written for the opposite polarization. We shall
show below that the observed behavior of the OS cannot be explained assuming
the initial state $|\psi \rangle $ to be of non-interacting electrons.
Therefore, we employ a model of interacting electrons, assuming that the
initial state close to $\nu =1$ and low enough temperature is a \textit{%
dilute gas of Skyrmions and anti-Skyrmions} and the OS is the sum of their
contributions. Following Ref. \cite{Fertig94} we write the initial state at
zero temperature at $\nu =1$ as $|\psi _{0}\rangle =\prod_{m=0}^{N_{\phi
}-1}c_{m,\uparrow }|0\rangle $, and the wave function of a single Skyrmion
or anti-Skyrmion as

\begin{eqnarray}
|\psi _{-}\rangle &=&\prod_{m=0}^{N_{\phi }-1}(u_{m}c_{m,\downarrow
}^{\dagger }+v_{m}c_{m+1,\uparrow }^{\dagger })|0\rangle ,  \label{states} \\
|\psi _{+}\rangle &=&\prod_{m=0}^{N_{\phi }-1}(-u_{m}c_{m+1,\downarrow
}^{\dagger }+v_{m}c_{m,\uparrow }^{\dagger })c_{0,\downarrow }^{\dagger
}|0\rangle ,  \notag
\end{eqnarray}%
with $|u_{m}|^{2}+|v_{m}|^{2}=1$ and $u_{m}\rightarrow 0$ as $m\rightarrow
\infty $. The free electron model is obtained by setting $u_{m}=0$ and $%
v_{m}=1$ for all $m$.

The OS is proportional to the absolute value squared of the matrix elements $%
M=\langle i|\sum_{m,s}h_{m,s}^{\dagger }c_{m,s}^{\dagger }|f\rangle $, and
it is straightforward to show that it can be written explicitly for the four
lines as

\begin{eqnarray}
|M_{X_{\downarrow ,\uparrow }}|^{2} &=&N_{a}\left( |v_{0}|^{2}+\sum
|u_{m-1}|^{2}|v_{m}|^{2}\right) +  \notag \\
&+&N_{s}\left( \sum |v_{m-1}|^{2}|u_{m}|^{2}\right) ,  \label{M} \\
|M_{T_{\downarrow }}|^{2} &=&N_{\phi }-(N_{s}+N_{a})\left( 1+\sum {%
|u_{m-1}|^{2}|v_{m}|^{2}+|u_{m}|^{2}}\right) ,  \notag \\
|M_{T_{\uparrow }}|^{2} &=&(N_{s}+N_{a})\left( |u_{0}|^{2}+\sum
|u_{m-1}|^{2}|u_{m}|^{2}\right) ,  \notag
\end{eqnarray}%
where all sums are over $m$ from $m=1$ to $m=N_{\phi }-1$, and $N_{s}$ and $%
N_{a}$ are the number of Skyrmions and anti-Skyrmions, respectively.
Equation (3) allows us to calculate the OS dependence on $\nu $ at zero
temperature, where $N_{s}=N_{\phi }(\nu -1)$,$\;N_{a}=0$ for $\nu >1$ and $%
N_{s}=0,\;N_{a}=N_{\phi }(1-\nu )$ for $\nu <1$. It is easy to see
that the free electron model does not yield the observed behavior
of the trion OS. This model predicts $|M_{T_{\uparrow }}|^{2}=0$
for any $\nu $, in a clear contradiction with the experimental result, which shows a strong $%
T_{\uparrow }$-absorption away from $\nu =1$. Furthermore, it gives a
constant $T_{\downarrow }$-absorption for $\nu <1$ when changing $\nu $ at
constant $n_{e}$ ($B$-scan), while the experiment shows a sharp drop [Fig.
3(a)]. In the Skyrmionic model, on the other hand, the sharp maximum in the $%
T_{\downarrow }$-absorption and the minimum in $T_{\uparrow }$ are readily
explained assuming that at least $u_{0}\neq 0$. According to the model, as $%
\nu $ is tuned away from 1, the Skyrmion effects reduce the $T_{\downarrow }$%
-absorption and transfer the corresponding OS to the $T_{\uparrow }$%
-absorption.

The behavior of the $X$-absorption, which vanishes for $\nu >1$, sets an
upper bound for the Skyrmion size. It follows from Eq. (3) that the $X$%
-absorption is strongly asymmetric with respect to $\nu =1$. In
order to explain the observed vanishing of the OS for $\nu >1$ one
has to assume that $u_{m}\approx 0$ for $m\geq 1$. This implies
that only electrons at $m=0$ and $1$ contribute significantly to
the Skyrmion. This result follows from the assumption that the
exciton is formed in orbitals where both spin states are empty,
implicitly allowing the nearest neighboring orbitals to be
occupied [Eq. (1)]. If one requires that these orbitals are also
unoccupied, the upper bound on the Skyrmion size will be higher:
in order for $|M_{X}|^{2}$ to vanish for $\nu >1$ it will suffice to assume that all $u_{m}\approx 0$ for $%
m\geq m_{0}$, where $m_{0}$ is the number of neighbors considered. Thus, the
actual number of electrons involved in a Skyrmion may be larger than two,
but of this order of magnitude. The conclusion is that only a few electron
orbitals participate in the Skyrmion texture.

Figure 3(b) depicts the calculated OS of $T\downarrow$ and
$T\uparrow$ at zero temperature for both the free electron model
and for a small Skyrmion ($u_{m\geq 1}=0$). It is seen that the
calculation reproduces qualitatively the experimental data. The
asymmetry around $\nu =1$ is due to the change in $N_{\phi }$ as
$B$ is scanned. The effect of a finite temperature is particularly
important close to $\nu =1$, where thermal spin excitations are
dominant. As a result, in the experiment the respective peak and dip in the OS of $T_{\downarrow }$ and $%
T_{\uparrow }$ are smoothed.

\begin{figure}[tbp]
\vspace*{3mm} {\hbox{\epsfxsize=83mm \epsffile{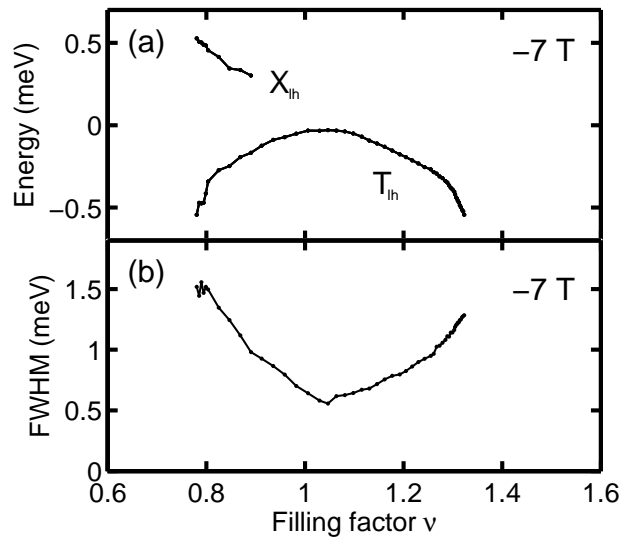} }}
\caption{ (a) The energy of the $X_{lh,\downarrow }$ and
$T_{lh,\downarrow }$ as a function of $n_{e}$ at constant $B$. The
scale was shifted by 1530.3 meV for clarity. (b) FWHM of the
$T_{lh,\downarrow }$. Notice the symmetry of the energy and width
around $\protect\nu =1$.} \label{fig4}
\end{figure}

Remarkably, the symmetry of the $T$-absorption around $\nu =1$ is observed
in the peak energy and width as well. Figure 4(a) shows the energies of the $%
X_{lh,\downarrow }$ and $T_{lh,\downarrow }$ as a function of $\nu $. It is
seen that the $T_{lh}$ energy is maximal at $\nu =1$ and drops symmetrically
at both sides. Similarly, we show in Fig. 4(b) the full width at half
maximum of the $T_{lh}$ peak in the same polarization. It is clearly seen
that the width is minimal at $\nu =1$ and rises symmetrically at both sides.
These observations are another manifestation of the electron-hole symmetry
of the Skyrmionic ground state \cite{Fertig94}. A quantitative treatment of
this behavior is, however, beyond the scope of this paper.

Finally, we wish to note that $X-T$ splitting is also observed at
higher Landau levels [see Fig. 1(a) around $\pm 2.5$ T]. This
splitting occurs around $\nu =3$, and another one is observed
around $\nu =5$. This might provide an opportunity to explore the
existence of Skyrmions at high filling factors as well
\cite{Song99}.

This work was supported by the Israel Academy of Science. J.G wishes to
acknowledge fruitful discussions with E. Mariani.




\begin{thebibliography}{99}
\bibitem{CommentTriplet} At low densities we observe a splitting of the peak
into exciton and triplet trion. 
\bibitem{Comment-ne} Each spectrum was normalized by dividing by the
PC efficiency, which depends on the gate voltage. The PC
efficiency was determined at high photon energy (1539 meV) at
$B=0$.
\end{thebibliography}
\end{document}